\documentclass[nohyperref]{article}

\usepackage{microtype}
\usepackage{graphicx}
\usepackage{subfigure}
\usepackage{booktabs} 
\usepackage{slashbox}

\usepackage[accepted]{icml2022}
\usepackage{hyperref}
\usepackage{balance}

\usepackage{mathtools} 
\usepackage[accepted]{icml2022}

\usepackage{amsmath}
\usepackage{amssymb}
\usepackage{mathtools}
\usepackage{amsthm}

\usepackage{paralist}
\usepackage[inline]{enumitem}

\usepackage{soul}
\usepackage{color}
\usepackage{wallpaper}
\usepackage{graphicx}
\usepackage{numprint}
\usepackage{paralist}
\usepackage{xspace}
\usepackage{xcolor}

\usepackage[some]{background}
\usepackage[11pt]{moresize}
\usepackage{t1enc}

\definecolor{orangeN}{rgb}{1,.5,0}
\definecolor{blueN}{rgb}{.2, .59, .88}
\definecolor{purpleN}{rgb}{.294118, 0, .509804}
\definecolor{greenN}{rgb}{.421, .578, .241}
\definecolor{pinkN}{cmyk}{0, 0.7808, 0.4429, 0.1412}
\definecolor{grayN}{gray}{0.6}

\newcommand{\mila}[1] {{\color{purpleN} \bf{[mila here]}}} 
 
\newcommand{\exvo}{\textsc{ExVo}}
\newcommand{\exmulti}{\textsc{ExVo-MultiTask}}
\newcommand{\exgen}{\textsc{ExVo-Generate}}
\newcommand{\exfew}{\textsc{ExVo-FewShot}}
\newcommand{\humevb}{\textsc{Hume-VB}}

\newcommand{\ds}{\textsc{DeepSpectrum}}
\newcommand{\cmp}{\textsc{ComParE}}
\newcommand{\opensmile}{\textsc{openSMILE}}
\newcommand{\openxbow}{\textsc{openXBOW}}
\newcommand{\egm}{\textsc{eGeMAPS}}

\newcommand{\boaw}{\textsc{BoAW}}

\newcommand{\ie}{i.\,e., }

\newcommand{\cf}{{cf.\ }}

\usepackage[capitalize,noabbrev]{cleveref}

\theoremstyle{plain}

\theoremstyle{definition}

\theoremstyle{remark}

\usepackage[textsize=tiny]{todonotes}

\icmltitlerunning{The ICML 2022 Expressive Vocalizations Workshop and Competition}

\begin{document}

\twocolumn[
\icmltitle{The ICML 2022 Expressive Vocalizations Workshop and Competition: \\ Recognizing, Generating, and Personalizing Vocal Bursts}


\icmlsetsymbol{equal}{*}

\begin{icmlauthorlist}

\icmlauthor{Alice Baird}{hume}
\icmlauthor{Panagiotis Tzirakis}{hume}
\icmlauthor{Gauthier Gidel}{mila,cifar}
\icmlauthor{Marco Jiralerspong}{mila}
\icmlauthor{Eilif B. Muller}{mila,cifar}
\icmlauthor{Kory Mathewson}{deep}
\icmlauthor{Björn Schuller}{imp}
\icmlauthor{Erik Cambria}{sent}
\icmlauthor{Dacher Keltner}{hume,berk}
\icmlauthor{Alan Cowen}{hume}
\end{icmlauthorlist}

\icmlaffiliation{hume}{Hume AI, USA}
\icmlaffiliation{deep}{DeepMind, Canada}
\icmlaffiliation{mila}{Mila \& University of Montreal, Canada}
\icmlaffiliation{cifar}{Canada Cifar AI Chair, Canada}
\icmlaffiliation{imp}{Imperial College London, UK}
\icmlaffiliation{sent}{Nanyang Technological University, Singapore}
\icmlaffiliation{berk}{University of California Berkeley, USA}

\icmlcorrespondingauthor{Alice Baird}{alice@hume.ai}

\icmlkeywords{affective computing, multi-task learning, few-shot learning, audio generation, generative adversarial networks}

\vskip 0.3in
]



\printAffiliationsAndNotice{}  

\begin{abstract}
The ICML \textbf{Ex}pressive \textbf{Vo}calization (\exvo{}) Competition is focused on understanding and generating vocal bursts: laughs, gasps, cries, and other non-verbal vocalizations that are central to emotional expression and communication. \exvo{} 2022 includes three competition tracks using a large-scale dataset of 59\,201 vocalizations from 1\,702 speakers. The first, \exmulti{}, requires participants to train a multi-task model to recognize expressed emotions and demographic traits from vocal bursts. The second, \exgen{}, requires participants to train a generative model that produces vocal bursts conveying ten different emotions. The third, \exfew{}, requires participants to leverage few-shot learning incorporating speaker identity to train a model for the recognition of 10 emotions conveyed by vocal bursts. This paper describes the three tracks and provides performance measures for baseline models using state-of-the-art machine learning strategies. The baseline for each track is as follows, for \exmulti{}, a combined score, computing the harmonic mean of Concordance Correlation Coefficient (CCC), Unweighted Average Recall (UAR), and inverted Mean Absolute Error (MAE) ($S_{MTL}$) is at best, 0.335 $S_{MTL}$; 
for \exgen{}, we report Fr\'{e}chet inception distance (FID) scores ranging from 4.81 to 8.27 (depending on the emotion) between the training set and generated samples. We then combine the inverted FID with perceptual ratings of the generated samples ($S_{Gen}$) and obtain 0.174 $S_{Gen}$; and for \exfew{}, a mean CCC of 0.444 is obtained. 
\end{abstract}

\section{Introduction} 

The \textbf{Ex}pressive \textbf{Vo}calizations Workshop and Competition (\exvo) is a novel competition-based workshop that addresses a critically overlooked problem in computer audition: understanding and generating vocal bursts, non-linguistic vocalizations that are central to the expression of emotion and to human communication more generally. Given that vocal bursts have largely been overlooked in the field of machine learning, likely due to the scarcity of vocal burst data, this first iteration of \exvo{} introduces a large-scale dataset of human vocalizations as a means to explore a wide range of computational approaches for understanding vocal bursts. In the present paper, we describe competition data, detail the three sub-challenges of the competition, and, to accelerate progress by competition teams, provide baseline performance benchmarks using state-of-the-art methods. 

The dataset used within the \exvo{} competition, the Hume Vocal Bursts dataset (\humevb) comprises 59,201 recordings totaling more than 36 hours of audio data from 1\,702 speakers. To our knowledge, this is substantially larger than than any previously available dataset of human vocal bursts. The recordings in \humevb{} are rich and diverse in a number of ways that present unique opportunities for understanding vocal bursts: \begin{inparaenum}[(1)] \item They were self-recorded by speakers in 4 countries spanning 3 native languages, the USA (English), China (Mandarin), South Africa (English), and Venezuela (Spanish). \item From an audio quality perspective, the recordings are diverse in acoustic conditions and noise profiles, as they were collected within speakers' homes via their own microphones, and can be considered `in-the-wild' (capturing uncontrolled and realistic settings). \item The breadth of vocal bursts represented in the dataset is unprecedented, capturing 10 classes of emotional expression each rated for intensity on a [0:\;100] scale. Thus, all three sub-challenges can be approached using multi-label regression methods, although separate modeling of each class is also permitted\end{inparaenum}.

These properties of the \humevb{} dataset enable the modeling of granular characteristics of vocal bursts, moving beyond the coarse classification of laughter~\cite{truong2005automatic} or cries~\cite{tuduce2018my}. Indeed, in-the-wild vocal bursts defy such coarse classification in two important ways: first, a single vocal burst often blends together distinct classes such as laughs, cries, and gasps; and second, two vocal bursts that belong to the same class, such as laughter, can have distinct meanings, such as amusement or embarrassment ~\cite{cowen2019mapping}. Thus, for the first time in a machine learning context, the \humevb{} dataset will enable participants to model continuous blends of utterances such as laughs, cries, and gasps as well as the distinct meanings of different laughs (amusement, awkwardness, and triumph), cries (distress, horror, and sadness), gasps (awe, excitement, fear, and surprise), and more.

With these goals in mind, and given the novelty and importance of modeling vocal bursts in multiple sub-fields of machine learning, \exvo{} 2022, explores three machine learning tasks. The first, \exmulti, requires participants to train a multi-task predictive model to jointly learn to measure the 10 continuous dimensions of emotional expression along with speaker age and native-country. Second, \exgen{} requires participants to train a generative model that produces vocal bursts conveying each of the 10 dimensions of emotional expression. Finally, \exfew{} requires participants to train a predictive model to measure the 10 dimensions of emotional expression while incorporating speaker identity information using two-shot speaker personalization. 

In this paper, we describe the \humevb{} dataset in detail (\Cref{sec:data}), provide rules for the three tasks (\Cref{sec:tasks}), and present baseline methods for each task (\Cref{sec:baselines}). We summarize our results in \Cref{sec:results} and conclude with a discussion of insights from baseline development in \Cref{sec:conclusions}.

\section{The Competition Dataset}
\label{sec:data}
The \exvo{} competition relies on \humevb{}, a large-scale dataset of emotional non-verbal vocalizations (vocal bursts). This dataset consists of 36\,:47\,:04  (HH\,:MM\,:SS) of total audio data from 1\,702 speakers, aged from 20 to 39 years old. The data was gathered in 4 countries with broadly differing cultures: the United States, China, South Africa, and Venezuela. Furthermore, the data is `in-the-wild', recorded at home via the speakers own microphones.

Each vocal burst has been labeled in terms of the intensity of ten different expressed emotions, each on a [1\,:100] scale, and these are averaged over an average of 85.2 rater’s responses\begin{inparaenum}[,]\item \textit{Amusement}\item \textit{Awe}\item \textit{Awkwardness}\item \textit{Distress}\item \textit{Excitement}\item \textit{Fear}\item \textit{Horror}\item \textit{Sadness}\item \textit{Surprise}\item \textit{Triumph}.\end{inparaenum} 

Participants within \humevb were recruited via a range of crowdsourcing platforms (Amazon Mechanical Turk, Clickworker, Prolific, Microworkers, and RapidWorker). Participants heard a `seed' vocal burst and were instructed to use their computer microphone to record themselves mimicking the vocal burst such that their imitation would be perceived to convey similar emotions to the original recording. Participants completed 30 trials per survey and could complete multiple versions of the survey. All participants provided informed consent and all aspects of the study design and procedure were approved by Heartland IRB.

\begin{figure}
    \centering
    \includegraphics[width=0.8\columnwidth]{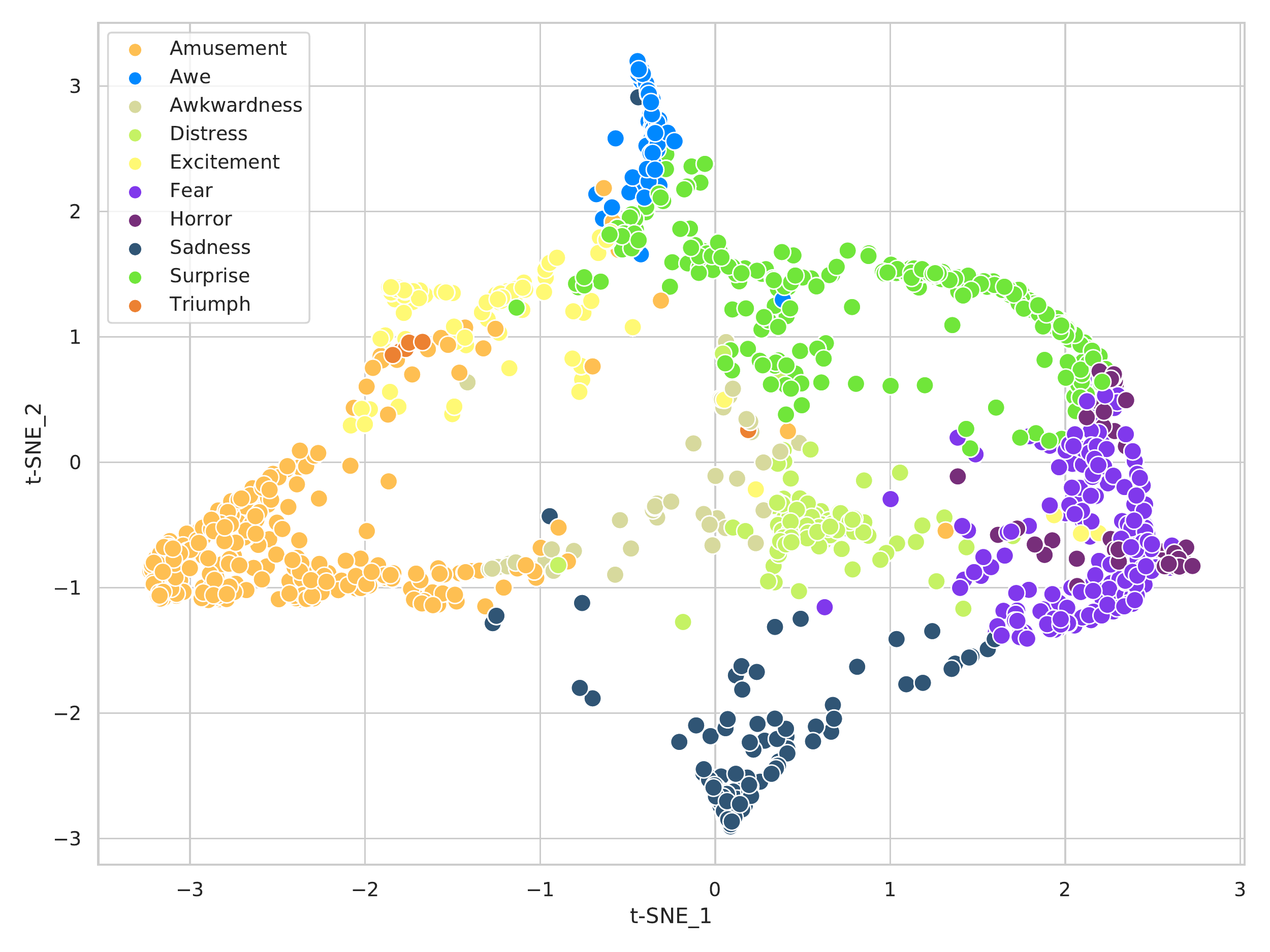}
    \vspace{-0.5cm}
\caption{A t-SNE representation of the emotional expression (normalized intensity scores) across the \humevb{} training set for each of the ten classes, used in all three \exvo{} tasks.}
    \label{fig:tsne}
    \vspace{-0.6cm}
\end{figure}

In \Cref{fig:tsne}, the distribution of emotional expressions, based on the intensity scores across the training set is visualized using t-SNE. We can see that the expressions vary continuously, with clearly defined regions corresponding to each expressed emotion as well as continuous gradients between emotions (e.\,g., amusement and excitement). Of note, there are fewer samples that differentially convey \textit{Triumph}, so we expect this class to be more challenging to model.

The intensity ratings for each emotion were normalized to range from [0\,:1]. For our baseline experiments, the audio files were normalized to -3 decibels and converted to 16\,kHz, 16\,bit, mono (we also provide participants with the raw unprocessed audio, which was captured at 48\,kHz). No other processing was applied to the files. Thus, data processing strategies for speech enhancement may be beneficial. The data was subsequently partitioned into training, validation, and test splits, considering speaker independence and balance across classes of interest. 
In \Cref{tab:splits}, we tabulate the number of samples and speakers by native-country and gender for each split.

\begin{table}
\small
\centering
\caption{An overview of the \humevb{} data. Including (No.) Samples, Duration (HH:\,MM:\,SS), Speakers, (M)ale:(F)emale, and per native-country. The age range for speakers is 20:39.5 years. For the purposes of the competition, the test set is blinded.}
\vspace{0.2cm}
\resizebox{\columnwidth}{!}{ 

\begin{tabular}{l | r r r r}
\toprule
             & \textbf{Train}    & \textbf{Validation}      & \textbf{Test}     & $\sum$ \\
             \midrule
\textbf{HH:\,MM:\,SS}     & 12\,:19\,:06 & 12\,:05\,:45 & 12\,:22\,:12 & 36\,:47\,:04            \\
\textbf{No.}      & 19\,990   & 19\,396  & 19\,815   & 59\,201              \\
\midrule
\textbf{Speakers}     &   571    &   568    &   563   & 1\,702             \\
\textbf{F:M}          & 305:266  & 324:244  &   ---    & ---             \\
\textbf{USA}          &   206    &   206    &   ---    & ---           \\
\textbf{China}        &   79     &   76     &   ---    & ---          \\
\textbf{South Africa} &   244    &   244    &   ---    & ---        \\
\textbf{Venezuela}    &   42     &   42     &   ---    & ---            \\
\bottomrule
\end{tabular}
}
    \vspace{-0.7cm}
\label{tab:splits}
\end{table}

\section{The Competition Tasks}
\label{sec:tasks}
The three tasks of the \exvo{} competition explore several machine learning approaches to modeling vocal bursts. 

\subsection{\exvo{} Multi-Task Learning Track}

In the \exmulti{} track, participating teams are tasked with training multi-task models that use vocal bursts to jointly predict the expression of 10 emotions along with the age and native-country of the speaker. Continuous intensity scores are provided for each of the 10 emotion classes, which can thus be predicted using multi-label regression methods. The mean Concordance Correlation Coefficient (CCC)~(\Cref{eq:ccc}) should be reported. Age should be predicted using regression and Mean Absolute Error (MAE) should be reported. Native-country should be predicted using classification and Unweighted Average Recall (UAR) should be reported. Each team should provide predictions for all samples in the test set. The three performance metrics will be averaged using a harmonic mean (with MAE inverted) to determine the final standings.

\subsection{\exvo{} Emotion Generation Track}

In the \exgen{} track, participating teams are tasked with training generative models that produce vocal bursts that differentially convey each of the 10 emotions. Each team will submit 100 generated samples for each emotion class they have chosen to target. The vocal bursts will be evaluated using both automated~(\Cref{eq:fid}) and human evaluation methods (\Cref{eq:heep}). The participants should report the Fréchet inception distance (FID)~\cite{heusel2017gans} between generated samples and the training set, \cf \Cref{sec:eval}. In place of using Inception features, as in the original FID score, we will use features from a classifier trained on the training set of \humevb. 

After each participating team submits their generated samples, the organizers will collect human ratings of five randomly (without replacement) drawn samples from the 100 given for each emotion. The ratings for the five samples will be obtained from the same participant population that contributed to the \humevb{} dataset. These ratings will be compared to the targeted emotions for each sample. Specifically, the Pearson correlation coefficient will be computed between normed (0-1) average intensity ratings for the 10 classes and the identity matrix consisting of dummy variables for each class. 

\subsection{\exvo{} Few-Shot Emotion Recognition Track}

In the \exfew{} track, participating teams are tasked with performing personalized recognition of emotional vocalizations using few-shot learning. Personalization is achieved by incorporating speaker identity~(permitting the model to learn factors such as the pitch, intensity, and fundamental frequency of the speaker's voice). To the best of our knowledge, this is the first challenge to tackle personalized recognition of vocal bursts. 

In this track, two labeled samples per speaker are to be used as the `support set' for personalization (2-shot learning). Identity labels are provided for all samples in the training (support) and evaluation (query) sets. A week before the submission deadline for the \exfew{} sub-challenge, two labeled samples will be provided for each speaker in the test set. As in \exmulti{}, participants will then use their models to provide personalized predictions of $10$ expressed emotions for each sample in the test set. Performance will be evaluated using the CCC~(\Cref{eq:ccc}).

\subsection{General Guidelines}
To participate in the \exvo{} 2022 competition, all participants are asked to provide a completed copy of the Hume-VB End-User License Agreement (EULA) (more details can be found on the competition homepage\footnote{\url{http://competitions.hume.ai/exvo2022}}). In addition, participants should submit a paper describing their methods and results that meets the official ICML guidelines. (The \exvo{} workshop is also accepting contributions on related topics.) To obtain test scores, participants should submit their test set predictions or generated samples to the competition organizers (each team can do this up to 5 times) with final scores being computed using the most recent submission). Participants are free to compete in any or all of the tasks. We also ask participants to upload their code along with their paper submission, and the organizers will reproduce and evaluate selected submissions. 


\section{Baseline Methods}

For the baseline of each track in the \exvo{} competition, we apply modeling strategies using established methods known to perform well for audio-based machine learning. 


\underline{\exmulti}: We train a baseline model for this task using common acoustic features extracted from the vocal bursts (made available to participants). We extract several sets of features that have been successfully deployed for related tasks~\cite{Schuller13-TI2,Schmitt16-ATB,baird2019using}. One feature vector is extracted per sample for each feature set. 

Using the \opensmile~toolkit, we extracted the 6,373-dimensional \cmp~set and the 88-dimensional \egm~ set. The 2016 COMputational PARalinguistics ChallengE (\cmp)~\cite{schuller2016INTERSPEECH} set contains 6,373 static features computed based on functionals from low-level descriptors (LLDs)~\cite{Eyben13-RDI,Schuller13-TI2}. The extended Geneva Minimalistic Acoustic Parameter Set (\egm)~\cite{eyben2015geneva}, which is smaller in size (88-dimensions), was designed for affective-based computational paralinguistic tasks. We also computed a Bag-of-Audio-Words (\boaw) representation from the \cmp~LLD's, utilizing the \openxbow~ toolkit~\cite{Schmitt17-OIT}. For our baseline, with \boaw{} we computed codebooks of size 125, 250, 500, 1\,000, and 2\,000. Finally, we extracted the 4,096-dimensional feature set of deep data representations using the \ds~ toolkit~\cite{Amiriparian17-SSC}. In this contribution, we utilize the default parameters from the \ds~ authors, utilizing VGG-19~\cite{simonyan2014very} pre-trained on ImageNet~\cite{russakovsky2015imagenet}, and the viridis colormap for the spectrograms. 

The baseline model for \exmulti{} applies a multi-task learning (MTL) network with hard-parameter sharing, in other words, the three tasks share the first two fully-connected layers which then feed into task-specific output layers~\cite{ruder2017overview}. In more detail, the MTL network consists of fully-connected layers, with layer normalization applied only between the first two shared layers. A leaky rectified linear unit (Leaky ReLU) is used as the activation function for all layers.

\begin{figure*}
    \centering
    \includegraphics[width=\linewidth]{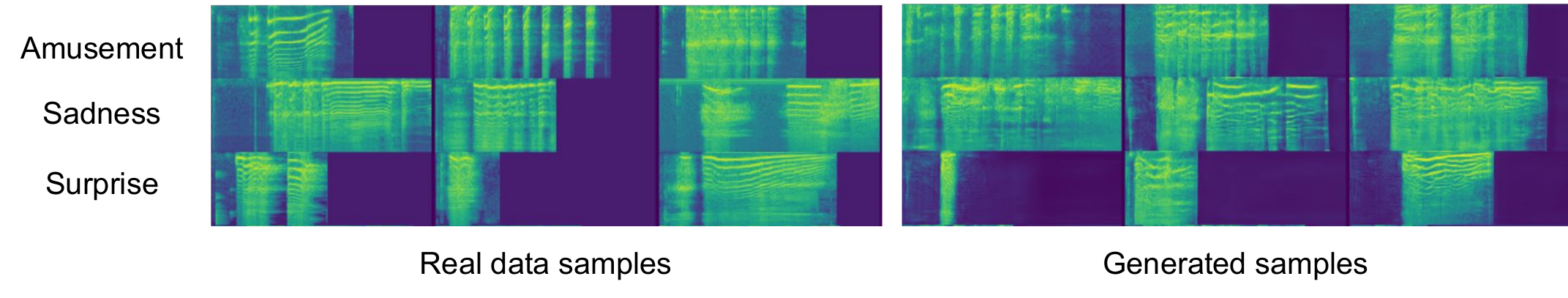}
    \vspace{-7mm}
    \caption{\small (Left) Mel spectrograms of nine original samples from the training set. (Right) nine samples generated by the \exgen{} baseline system. Each row corresponds to a different emotion, Amusement, Sadness, Surprise.
    For each spectrogram, the y-axis corresponds to the log-magnitude of each frequency and the x-axis corresponds to 4 seconds of audio.}
    \label{fig:spectrograms}
\end{figure*}

\underline{\exgen}: Our baseline generative model is based on the work by~\cite{afsar2021generating} which uses a Multi-Scale Gradient Generative Adversarial Network (MSG-GAN)~\cite{karnewar2020msg} architecture trained to generate Mel spectrograms of audio samples~\cite{kumar2019melgan}. The independent training bins for each class are created by selecting all samples with an emotional value of 1 for that class. The generated spectrograms are converted back into audio signals using the `fast' Griffin Lim inversion method implemented by the audio toolkit Librosa~\citep{brian_mcfee_libro}. We present a comparison of a selection of generated spectrograms against some real spectrograms from the \humevb{} dataset in~\Cref{fig:spectrograms}.
\footnote{Selection of generated samples, May 3, 2022: \url{https://bit.ly/3LvQQa6}}.

\underline{\exfew}: Our baseline for this task integrates a few-shot learning approach with an end-to-end convolutional recurrent neural network architecture~\cite{tzirakis2017speech}. In particular, our end-to-end architecture is comprised of three blocks of 1-D CNN layers with a Leaky ReLU (with negative slope equal to $0.01$) activation function and max-pooling operations. 
Both convolution and pooling operations are performed in the time domain, i.\,e., using the raw waveform as input. Based on previous studies~\cite{tzirakis2017speech, tzirakis2018end2you}, we perform convolution with a small kernel size ($8,6,6$), and stride of one, and a large kernel ($10,8,8$) and stride size for the max-pooling.
After the convolution network, a 2-layer LSTM is added to extract temporal characteristics in the signal. 

The training of our model is performed without taking into consideration the speaker's identity information. As such, we use the training set to train our model and the best performing model on the validation set is selected for inference. In this manner, the model will learn to recognize vocal bursts independent of the speaker. We leave a more suited training process to the participants (e.\,g., episodic training). On inference, however, we consider the speaker's identity. In more detail, given the 2 samples in the support set per speaker, we fine-tune the linear and the recurrent network using these samples before computing the prediction of the samples in the query set.

\section{Baseline Experiments}
\label{sec:baselines}


We adopted commonly used strategies to produce baseline scores for each task. We provide reproducible code supporting each baseline model on GitHub\footnote{\url{http://github.com/HumeAI/competitions/tree/main/ExVo2022}}.

\subsection{Experimental Setup}

\underline{\exmulti}. For training the multi-task model, the Adam optimization method is used~\cite{adam}, and we sum the  three losses calculated for each target: (i) the Mean Squared Error (MSE) for the emotion recognition and the age detection task, and (iii) the cross-entropy loss for the native-country detection. After tuning the hyperparameters on the validation set, we find that a learning rate ($lr$) of $10^{-3}$ and batch size ($bs$) of $8$ is optimal. To avoid overfitting, we apply an early stopping strategy with a patience of $5$ epochs, and a maximum number of epochs of $20$. 

\underline{\exgen}. We use a similar hyper-parameter configuration to~\citealp{afsar2021generating}, except that some models were trained with the ExtraAdam optimizer~\cite{gidel2018variational} and we use different learning rates for the generator ($3\times10^{-3}$) and the discriminator ($3\times10^{-4}$). For each emotion, we trained our model on the train and validation set for 1\,450 to 5\,000 epochs (which corresponds to 72 hours of training on an RTX 8\,000).

\underline{\exfew}. For training the model for this task, we used the MSE loss function and the Adam~\cite{adam} optimization algorithm with initial learning rate of $10^{-4}$. We used a mini-batch of $8$ samples, and the weights of the network have been initialized with Kaiming uniform~\cite{he2015delving} initialization with the biases to be set to zero.

\subsection{Evaluation Metrics}
\label{sec:eval}

Different evaluation metrics are reported for each task. For \exmulti, we use the mean Concordance Correlation Coefficient (CCC) of all emotions ($\mathcal{\hat{C}}$) for the emotion recognition task, the Mean Absolute Error (MAE), which is then inverted as $1/MAE$ ($\mathcal{\hat{M}}$), for the age detection task, and the Unweighted Average Recall (UAR) ($\mathcal{\hat{U}}$) for the native-country classification. All of the metric values are combined to an overall metric score using the harmonic mean. For \exfew, we also report the mean CCC of all emotions. In more detail,

\begin{itemize}
    \item \textit{Concordance Correlation Coefficient (CCC).} Between the predictions ($x$) and the ground truth ($y$) of an emotion dimension ($e$), the CCC function is defined as

    \begin{equation}
        \begin{split}
            \mathrm{\mathcal{C}_e} = \dfrac{2 \sigma_{xy}^2}{\sigma_x^2 + \sigma_y^2 + (\mu_x - \mu_y)^2},
        \end{split}
    \label{eq:ccc}
    \end{equation}

    \noindent where $\mu_{x} = \mathbb{E}(\mathbf x)$, ${\mu_{y} = \mathbb{E}(\mathbf y)}$, ${\sigma_x^2 = \mbox{var}(\mathbf x)}$, $\sigma_y^2 = \mbox{var}(\mathbf y)$, and $\sigma^2_{xy} = \mbox{cov}(\mathbf x, \mathbf y)$. As our emotion recognition task provides ten emotions, the overall CCC is defined as the mean for each emotion dimension: $\mathcal{\hat{C}} = \sum_{i=1}^{10} \mathcal{C}_i/10$.

    \item \textit{Harmonic Mean.} The harmonic-mean of all scores ($S_{MTL}$) is defined as

    \begin{equation}
        \mathrm{S_{MTL}} = \frac{3}{(1/\mathcal{\hat{C}} + 1/\mathcal{\hat{M}} + 1/\mathcal{\hat{U}})}. 
    \end{equation}
    
\end{itemize}

For \exgen, we offer two core strategies for evaluating the generated samples and combine these as our baseline. Principally, the Fr\'{e}chet inception distance (FID)~\citep{heusel2017gans}, which will be calculated using $1,000$ generated samples against the training samples. It will be computed separately for generated samples targeting each emotional class. Competition standings will then be presented for each emotion separately as well as for the full distribution. It is expected that some participating teams may elect to compete only for a subset of targeted emotions. In addition, human ratings will be gathered for a random subset of $5/100$ samples per targeted emotion per team. 

\begin{itemize}
    \item \textit{Fr\'{e}chet inception distance (FID)} to compare the distribution of samples that have been generated against the distribution of the original data, defined as 
    \begin{equation}
        \mathrm{FID} =\|\mu - \mu^*\|^2 + Tr( C + C^* - 2(C  C^*)^{1/2}), 
        \label{eq:fid}
    \end{equation}
    where $\mu$ and $C$ (resp. $\mu^*$ and $C^*$) correspond to the empirical mean and covariance of the generated distribution (resp.\ validation set) whose samples are first mapped to the feature space of a classifier trained on the \humevb{} data. 
    
    \item \textit{Human-Evaluated Expression Precision (HEEP)}. The evaluation metric computed from human ratings of generated vocal bursts will be defined as 
    \begin{equation}
        \mathrm{HEEP} = \sigma_{TH}/\sqrt{\sigma_T^2\sigma_H^2 },
        \label{eq:heep}
    \end{equation}
    where $T$ corresponds to the vectorized target matrix, a dummy matrix of size $N$ (number of generated vocal bursts) by 10 (emotions), with ones for targeted emotions and zeros for non-targeted emotions, and $H$ corresponds to the vectorized rating matrix, a matrix of size $N\times10$ with entries corresponding to the average human intensity ratings of each generated vocal burst.

     \item As a final metric for the \exgen{} baseline, we calculate $S_{GEN}$, we compute the mean between the inverted FID distance, and the HEEP score for each emotion ($e$); this is defined as 
    
        \begin{equation}
            S_{GEN}{_e}={\frac {1/FID_e+HEEP_e}{2}}.
        \end{equation}
    
\end{itemize}

\begin{table*}[]
\centering
\caption{Validation and baseline test scores for \exmulti. Reporting scores for the best seed on validation and standard deviation from 5 seeds. Reporting  mean CCC across the 10 (Emo)tional classes, UAR for the 4 class (Cou)ntry task (chance level 0.25 UAR ), and MAE for the Age regression task. Baseline is set as the harmonic mean between these metrics ($S_{MTL}$). Emphasised results indicate best scores, with the official baseline as Test $S_{MTL}$.   }
\vspace{0.1cm}
\resizebox{0.65\linewidth}{!}{ 
\begin{tabular}{l r |r r r c | c}
\toprule
            &           & \multicolumn{4}{c |}{\textbf{Validation}}                                                      & \textbf{Test}                                   \\
            & Dims.     & Emo-CCC  & Cou-UAR  & Age-MAE  & $S_{MTL}$                 & $S_{MTL}$     \\
\midrule
\cmp        & $6\,373$    & \textbf{0.416}   & \textbf{0.506}   & 4.222            & \textbf{0.349} $\pm$ 0.003  & \textbf{0.335} $\pm$ 0.002 \\
\egm        & $88$        &  0.353           & 0.423            & \textbf{4.011}   & 0.324 $\pm$ 0.005           & 0.314 $\pm$ 0.005          \\
\midrule
\boaw       &$125$        & 0.335            & 0.417            & 4.268            & 0.311 $\pm$ 0.004           & 0.299 $\pm$ 0.004          \\
            &$250$        & 0.354            & 0.423            & 4.197            & 0.319 $\pm$ 0.005           & 0.305 $\pm$ 0.004         \\
            &$500$        & 0.374            & 0.432            & 4.581            & 0.314 $\pm$ 0.002           & 0.302 $\pm$ 0.002          \\
            &$1\,000$     & 0.384            & 0.438            & 4.446            & 0.321 $\pm$ 0.004           & 0.307 $\pm$ 0.003          \\
            &$2\,000$     & 0.397            & 0.443            & 4.531            & 0.322 $\pm$ 0.002           & 0.303 $\pm$ 0.002          \\
\midrule
\textsc{DeepSpec.}         &  $4\,096$   &  0.369           & 0.456            & 4.413            & 0.322 $\pm$ 0.003           &0.305 $\pm$ 0.003         \\
\bottomrule
\end{tabular}}
\label{tab:results_multi}

\end{table*}

\section{Baseline Results} 
\label{sec:results} 


In the following, we discuss the baseline results obtained for each of the three tracks. 

\subsection{\exmulti}

Results for the \exmulti{} can be seen in \Cref{tab:results_multi}. Across each task, emotion, native-country, and age, all feature sets appear to be performing robustly. With a generally lower performance for the 4 class native-country task (best score on validation 0.51 UAR, chance level 0.25 UAR, with \cmp{} features) than would be typically possible from speech. These lower results for the native-country task, show that the cultural diversity in the set is less easily modeled from non-verbal vocalization with extracted features, indicating that there is strong similarity across cultures for these non-verbal expressions. For age, we see results which are within a state-of-the-art range (best score on validation 4.01 MAE in years with \egm{} features).  For the 10-class emotion regression task, we see a mean CCC of at best 0.42 on validation, utilizing the \cmp{} feature set; this indicates a moderate correlation across all the classes.

For specific features, we see the \cmp{} set performs well across all tasks, and sets the baseline for this task ($S_{MTL}$ 0.335). Interestingly, for \boaw{} features, we see a gradual increase in CCC and UAR as the codebook size increases. The inverse of this is seen for Age, with a decrease in MAE the larger the \boaw{} codebook size. Similarly, we also see a strong performance for age MAE with the smaller size \egm{} set, both indicating that the age task does not benefit from brute-force modeling, and a knowledge-based approach may be beneficial.

\begin{table*}[]
\centering
\caption{ (Above) Fr\'{e}chet inception distance (FID) between different emotions in training vs (val)idation sets and  (Below) the scores for samples generated for \exgen{} baseline. For each emotion, we evaluated 1,000 samples. The diagonal of the FID between train and validation set provides lower-bound on the best FID to expect from a generative model, going below this value may indicate that the generator merely memorized the training set. Across all emotions, the baseline for \exgen{} is $0.174$ $S_{Gen}$.}
\vspace{0.1cm}
\resizebox{0.75\linewidth}{!}{ 
\begin{tabular}{l | r r r r r r r r r r}
\toprule
  \backslashbox{\textbf{Train}}{\textbf{Val}}  
  & Amuse.                            & Awe                             & Awkward.                        & Distress                         & Excite.                         & Fear                            & Horror                         & Sadness                         & Surprise                         & Triumph                          \\
  \midrule
Amuse.                                                                                           & \textbf{0.634  } & 17.65                           & 8.40                            & 12.83                            & 10.78                           & 13.64                           & 13.61                          & 8.53                            & 15.37                            & 11.47                            \\ 
Awe                                                                                              & 17.65                             & \textbf{0.776} & 11.61                           & 8.34                             & 9.42                            & 13.15                           & 11.07                          & 14.60                           & 8.43                             & 16.23                            \\
Awkward.                                                                                         & 8.40                              & 11.61                           & \textbf{1.20 } & 3.73                             & 7.18                            & 5.81                            & 6.68                           & 7.90                            & 6.48                             & 9.80                             \\
Distress                                                                                         & 12.83                             & 8.34                            & 3.73                            & \textbf{0.866 } & 6.33                            & 4.22                            & 3.11                           & 8.43                            & 4.64                             & 10.88                            \\
Excite.                                                                                          & 10.78                             & 9.42                            & 7.18                            & 6.33                             & \textbf{0.697} & 8.43                            & 6.37                           & 10.34                           & 6.65                             & 10.97                            \\
Fear                                                                                             & 13.64                             & 13.15                           & 5.81                            & 4.22                             & 8.43                            & \textbf{0.649} & 2.88                           & 8.14                            & 4.10                             & 12.29                            \\
Horror                                                                                           & 13.61                             & 11.07                           & 6.68                            & 3.11                             & 6.37                            & 2.88                            & \textbf{1.25} & 9.62                            & 4.92                             & 12.58                            \\
Sadness                                                                                          & 8.53                              & 14.60                           & 7.90                            & 8.43                             & 10.34                           & 8.14                            & 9.62                           & \textbf{0.992} & 12.91                            & 16.58                            \\
Surprise                                                                                         & 15.37                             & 8.43                            & 6.48                            & 4.64                             & 6.65                            & 4.10                            & 4.92                           & 12.91                           & \textbf{0.341 } & 10.15                            \\
Triumph                                                                                          & 11.47                             & 16.23                           & 9.80                            & 10.88                            & 10.97                           & 12.29                           & 12.58                          & 16.58                           & 10.15                            & \textbf{3.76  } \\ 
\midrule
\multicolumn{11}{c}{\exgen{} Baseline (training vs generated samples)}       \\
\midrule
FID                                                                                              & 4.92                              & 4.81                            & 8.27                            & 6.11                             & 6.00                            & 5.71                            & 5.64                           & 5.00                            & 6.08                             & --                               \\
HEEP                                                                                             & 0.49                              & 0.46                            & 0.036                           & 0.32                             & 0.084                           & 0.042                           & 0.27                           & -0.033                          & 0.22                             & --                               \\
$S_{GEN}$                                                                                        & 0.347                             & 0.334                           & 0.078                           & 0.242                            & 0.125                           & 0.109                           & 0.224                          & 0.084                          & 0.192                            & 0.00                              

\\
\bottomrule
\end{tabular}}
\label{tab:fid_emotions}

\end{table*}

\begin{table*}[ht!] 
\caption{Test results (w.\,r.\,t. $\mathcal{\hat{C}}$) for the \exfew{} task. Reporting the individual CCC for each emotion and the average ($\mathcal{\hat{C}}$) across the 10 emotion dimensions when varying the number of hidden units ($\mathcal{H}$) $64$, $128$, and $256$. Baseline score is emphasized. } 
\label{tab:few_shot_results} 
\vspace{0.1cm}
\centering 
\resizebox{0.8\linewidth}{!}{ 
\begin{tabular}{l | r r r r r r r r r r | r}
\toprule
$\mathcal{H}$
  & Amuse. & Awe & Awkward. & Distress & Excite. & Fear & Horror & Sadness & Surprise & Triumph & $\mathcal{\hat{C}}$ \\ \midrule 
64 & 0.478 & 0.591 & 0.260 & 0.377 & 0.262 & 0.493 & 0.447 & 0.400 & 0.563 & 0.237 & $0.421 \pm 0.010$ \\
128 & 0.551 & 0.587 & 0.287 & 0.376 & 0.314 & 0.523 & 0.478 & 0.388 & 0.594 & 0.297 & $0.442 \pm 0.007$ \\
256 & 0.554 & 0.581 & 0.282 & 0.420 & 0.311 & 0.544 & 0.490 & 0.383 & 0.561 & 0.315 & $\textbf{0.444} \pm 0.006$ \\
\bottomrule 
\end{tabular} 
}
\vspace{-0.4cm}

\end{table*}

\subsection{\exgen} 
In our experiments, we trained a different generator on the samples of each individual emotion (\ie samples with value 1 for that emotion) except \emph{Triumph} (given the limited number of samples) as well as a generator on the data from all emotions simultaneously (note that we do not report the performance of this model, since the competition focuses on the generation of specific emotions). We restricted our training to US samples, but we evaluated our method using the FID between 1,000 generated samples and the validation set containing all the locations.
We observed that the training was considerably harder to stabilize when considering samples from all locations. While the use of ExtraAdam and different learning rates could somewhat stabilize the training of the model, restricting ourselves to just US samples was necessary to obtain meaningful results. We hypothesize that this difference may be due to US samples being on average of higher audio quality with less background noise (though further investigation is required to determine the exact cause).

There is a substantial margin for improvement by managing to train a generative model using all examples. Investigating conditional generative models to simultaneously solve all the generation tasks could also provide some improvements. Finally, we invite the participants to try other generative models such as, autoregressive models~\citep{Van16-WAV}, variational autoencoders~\citep{kim2021conditional,kong2020diffwave}, normalizing flows~\citep{esling2019flow}, or diffusion models~\citep{dhariwal2021diffusion}.

We present in Table~\ref{tab:fid_emotions} our FID metric across different emotions to give an idea of the order of magnitude of the metric on the \exvo{} dataset as well as the performance of our model. In \Cref{tab:fid_emotions} the baseline is given, where FID is calculated between training and generated samples. As well as this the HEEP score is provided and the baseline is given as the $S_{GEN}$ metric, see \Cref{sec:baselines}. Of interest we see that \textit{Amusement} and \textit{Awe} perform strongly in terms of FID and HEEP with more room for improvement across the other classes. Based on these current results, we also provide a selection of samples for the interested listener to evaluate.


\subsection{\exfew{}} 

In our experiments, we investigate how the performance of the model varies when different hidden units are used in the 2-layer LSTM. Essentially, we explore how the number of temporal dimensions affect the prediction of the ten emotion dimensions of each speaker. To this end, we experiment with $64$, $128$, and $256$ number of hidden units. Table~\ref{tab:few_shot_results} depicts the test results (w.\,r.\,t. $\mathcal{\hat{C}}$). Each experiment was run twice and the mean of each emotion is shown in the table, along with the mean (with the standard deviation) across all emotions which is the baseline of the task.

We observe that increasing the dimensionality improves the results with a $256$ hidden units to provide the best performing model, with mean CCC of $0.444\pm 0.006$. Using a size of $64$ is not adequate to fully capture the complex temporal dynamics in the signal, as the mean CCC is $0.421$ with higher standard deviation ($0.010$) compared to the other experiments. Finally, we see that using $128$ hidden units we get very similar mean results compared to using $256$ units, and as such we have not tried using higher number of hidden units (e.\,g. $512$).

\section{Conclusions} 
\label{sec:conclusions} 


With this contribution, we introduced the guidelines and baseline results for the first ICML Expressive Vocalizations (ExVo) competition. The competition investigates the understanding and generation of vocal bursts with a newly introduced large-scale and `in-the-wild' dataset, \humevb. In this year's competition, three tasks were introduced: (i) \exmulti{}, a multi-task learning track in which participants must jointly predict the native-country, age, and 10 emotion dimensions. We report a baseline score of \textbf{0.335 $S_{MTL}$ for \exmulti}, (ii) \exgen{}, in which participants should generate emotionally expressive vocal bursts. We report, an average \textbf{$S_{GEN}$ of 0.174 for \exgen{}} across all emotions, and for a single emotion (\textit{Amusement}) $0.347$ $S_{GEN}$. Currently, we do not report a score for Triumph, (iii) \exfew{} in which participants should personalize the prediction of 10 emotion dimensions in a few-shot manner. We report a baseline score of, \textbf{0.444 CCC for \exfew}. There are several aspects which can be explored by participants of the \exvo{} competition to improve on the provided baselines. Namely, via knowledge-based approaches targeted at fully harnessing the diversity present across the \humevb{} dataset, as well as modeling strategies suitable for the large quantity of data made available. 

\section*{Acknowledgments}


We thank Hume AI, Mila, and the National Film Board of Canada for their support in sponsoring this year's \exvo{} competition. Eilif Muller and Gauthier Gidel are supported by the Canada CIFAR AI Chairs Program.

\balance
\bibliography{main}
\bibliographystyle{icml2022}

\end{document}